\documentclass[twocolumn,showpacs,preprintnumbers,amsmath,amssymb,prl,aps,superscriptaddress,floatfix]{revtex4}
\usepackage{graphicx}
\usepackage{dcolumn}
\usepackage{bm}
\usepackage[dvips]{color}

\begin{document}

\title{Topological Winding and Unwinding in Metastable Bose-Einstein Condensates}

\author{Rina Kanamoto}
\affiliation{Department of Physics, University of Arizona, Tucson, AZ, 85721, USA}

\author{Lincoln~D. Carr}
\affiliation{Department of Physics, Colorado School of Mines, Golden, CO, 80401, USA}

\author{Masahito Ueda}
\affiliation{Department of Physics, Tokyo Institute of Technology, Meguro-ku, Tokyo 152-8551, Japan}
\affiliation{ERATO, JST, Bunkyo-ku, Tokyo 113-8656, Japan}

\date{\today}

\begin{abstract}
Topological winding and unwinding in a quasi-one-dimensional metastable
Bose-Einstein condensate are shown to be manipulated by
changing the strength of interaction or the frequency of rotation.
Exact diagonalization analysis reveals that quasidegenerate states
emerge spontaneously near the transition point, allowing a smooth crossover
between topologically distinct states.
On a mean-field level, the transition is accompanied by formation of grey
solitons, or density notches, which serve as an experimental signature of
this phenomenon.
\end{abstract}
\maketitle


Metastability of a physical system leads to a rich variety of
quantum phases and transport properties that are not present in
the ground state phase. An illustrative example is
superflow and phase slip in a narrow superconducting
channel~\cite{67:LA}. Other examples include Feshbach molecules
formed in high rotational states~\cite{07:Innsbruck} and
metastable quantum phases in higher Bloch bands in an optical
lattice~\cite{05:SS}. Recent experimental advances in cold
atoms/molecules have made it possible to realize excited,
metastable states which persist for a long time.  These states
provide an excellent medium in which to investigate fundamental
aspects of condensed matter systems such as topological
excitations and superfluidity~\cite{topological_excitations,99:L, superfluid,73:Bloch}.

It is widely believed that the angular momentum per particle in 
a weakly repulsive one-dimensional (1D) superfluid ring
system~\cite{superfluid, 73:Bloch} is quantized at $T=0$
and that there are discontinuous jumps between states having
different values of the phase winding number. In this
Letter, we point out that this applies only to the ground
state; continuous transitions do in fact occur between
metastable states of repulsive condensates. The underlying
physics behind this phenomenon is the emergence of a dark or
grey soliton train~\cite{00:CCR} which bifurcates from the
plane-wave solution and carries a fraction of the quantized
value of the angular momentum.

Starting with mean-field theory for scalar bosons subject to rotation,
we proceed through progressively deeper levels of insight into the quantum
many-body nature of this problem, making a link between semiclassical
and quantum solitons in metastable states.
We find that the phase slip, which allows a smooth crossover between topologically 
distinct states, is caused by a \emph{quantum soliton}.  The latter consists of a
linear superposition of the rotationally-invariant many-body eigenstates 
of the Hamiltonian~\cite{05:KSU}. In both Boguliubov theory and quantum
many-body theory the broken-symmetry soliton state is shown to be stable against perturbation.

This phenomenon can be realized by hot atoms confined in
fast-rotating circular waveguides or toroidal
traps~\cite{04:ringtrap}.  First, to obtain a metastable
uniform condensate one quickly stops the rotation and then
lowers the temperature. Second, one adiabatically changes the
angular frequency of the trap in the presence of a 
small arbitrary perturbation in the trapping potential.
This causes atoms to adiabatically take the higher-energy
path of a metastable soliton state, as we will show.  Third,
one stops the adiabatic change in the frequency at the correct
point to arrive at a different winding number. 
All of these processes can occur continuously. 


\begin{figure*}[t]
\includegraphics[scale=0.95]{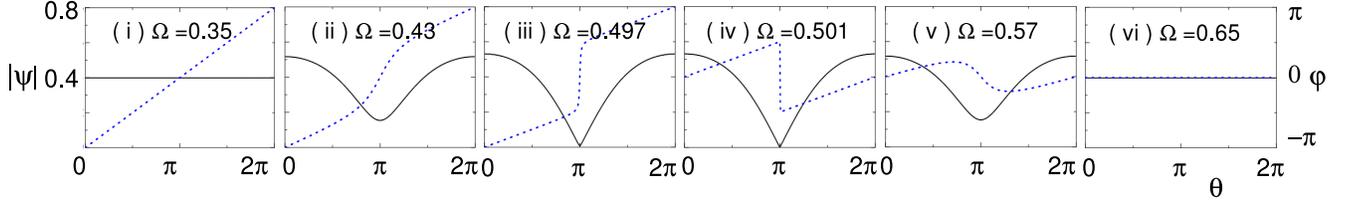}
\caption{
Amplitude (solid curves with the left reference), and
phase (dotted curves with the right reference) of metastable states
of the GPE for $g_{\rm 1D}N=0.6\pi$. Uniform solutions with different values
of the phase winding (i) $J=1$, and (vi) $J=0$ are smoothly connected
through the broken-symmetry grey soliton (ii)--(v)
with a self-induced phase slip at $\Omega=0.5$.
}
\label{fig1}
\end{figure*}

We consider a system of $N$ bosonic atoms in a quasi-1D torus
with radius $R$, under an external rotating drive with angular
frequency $2\Omega$. The length, angular momentum, and energy
are measured in units of $R$, $\hbar$, and $\hbar^2/(2 m
R^2)$, respectively. The Hamiltonian is given by the 
Lieb-Liniger Hamiltonian in a rotating frame of
reference~\cite{63:LL,footnote1}, 
\begin{eqnarray}
\hat{H}=\textstyle\int_0^{2\pi} d\theta
[\hat{\psi}^{\dagger}
(-i\partial_{\theta}-\Omega)^2\hat{\psi}
+g_{\rm 1D}\hat{\psi}^{\dagger 2}\hat{\psi}^2 /2],
\label{Hamiltonian}
\end{eqnarray}
where $g_{\rm 1D}$ characterizes the strength of the $s$-wave
interatomic collisions in 1D~\cite{98:MO} rescaled by
$\hbar^2/(2mR)$, $\theta$ is the azimuthal angle, and the
bosonic field operator satisfies periodic boundary
conditions: $\hat{\psi}(\theta)=\hat{\psi}(\theta+2\pi)$.
Since the Hamiltonian is periodic with respect to $\Omega$, the
properties of the system are periodic in $\Omega$ with period
1~\cite{footnote2}, in direct analogy to the reduced Brillouin
zone in a Bloch band~\cite{73:Bloch,06:BPSKCH}. Without loss of
generality we will henceforth restrict ourselves to $\Omega \in
[0,1)$.  The Hamiltonian is integrable via the boson-fermion
mapping in the Tonks-Girardeau (TG) limit $g_{\rm 1D} \gg
N$~\cite{60:G}, and via the Bethe ansatz in the
weak-interation limit $g_{\rm 1D}N \lesssim O(1)$ 
as well as intermediate-interaction regimes. 


We first show how continuous changes in the angular 
momentum occur in the weak-interaction regime for solutions of the 
Gross-Pitaevskii equation (GPE) 
$[(-i\partial_{\theta}-\Omega)^2+g_{\rm 1D}N|\psi(\theta)|^2]\psi(\theta)=\mu\psi(\theta)$, 
where $\psi$ is the order parameter normalized to unity
and $\varphi\equiv \mathrm{Arg}(\psi)$
is its phase. The single-valuedness of the wave function requires
$\varphi(\theta+2\pi)=\varphi(\theta)+2\pi J$, where $J\in \{0,\pm1,\pm2,\ldots\}$
is the topological winding number. 
Stationary solutions of the GPE for $g_{\rm 1D} \ge 0$ are either plane-wave
states $\psi(\theta)=e^{iJ\theta}/\sqrt{2\pi}$ or a grey soliton
train~\cite{00:CCR}
whose amplitude and phase are given by 
$|\psi(\theta)|=A \textstyle[1+\eta\ {\rm dn}^2(jK(\theta-\theta_0)/\pi, k)]^{1/2}$, and 
$\varphi(\theta)=\Omega\theta+B\ \Pi\left(\xi; jK(\theta-\theta_0)/\pi, k \right)$, respectively.
Here the amplitude $A\equiv \sqrt{K/[2\pi(K+\eta E)]}$; the phase pre-factor
$B\equiv({\cal S}/jK)
\sqrt{g_{\rm sn}h_{\rm sn}/2f_{\rm sn}}$; there are $j$ density notches in
the soliton train; $\eta=-2j^2K^2/{g_{\rm sn}}\in[-1,0]$ characterizes
the depth of each density notch; $k \in [0,1]$ is the elliptic modulus;
$K(k)$, $E(k)$, $\Pi(\xi,u,k)$, are
elliptic integrals of the first, second, and third kinds;
and dn$(u,k)$ is the Jacobi dn function.
The degeneracy parameter $\theta_0$ indicates
that the soliton solutions are broken-symmetry states.
We define
$f_{\rm sn} = \pi g_{\rm 1D}N/2 -2 j^2 K^2 +2 j^2 K E$,
$g_{\rm sn} = f_{\rm sn}+2 j^2 K^2$,
$h_{\rm sn} = f_{\rm sn}+2k^2 j^2 K^2$,
and ${\cal S}=1$ for $0 \leq \Omega < 0.5$,
${\cal S}=-1$ for $0.5\leq \Omega < 1$.
Then $\xi=-2(k j K)^2/f_{\rm sn}\leq 0$ , and only
when soliton solutions exist~\cite{k} is $k\neq 0$. 
In the limit $\eta \to -1$, $f_{\rm sn}$ approaches zero, and 
$g_{\rm sn}$ and $h_{\rm sn}$ both approach finite positive values; 
consequently, the wave function approaches the Jacobi sn function,
which corresponds to a dark soliton with a $\pi$-phase jump at $\theta_0$.
In the opposite limit $\eta \to 0$, both the amplitude and phase 
approach those of the plane-wave solution with the same phase 
winding $J$. These limiting behaviours
indicate that continuous change in topology of the condensate wave function is possible,
as illustrated in Fig.~\ref{fig1} (i)-(vi).
Henceforth, we consider the single soliton $j=1$ for simplicity,
but our discussion holds for arbitrary soliton trains $j>1$.

\begin{figure}
\includegraphics[scale=0.48]{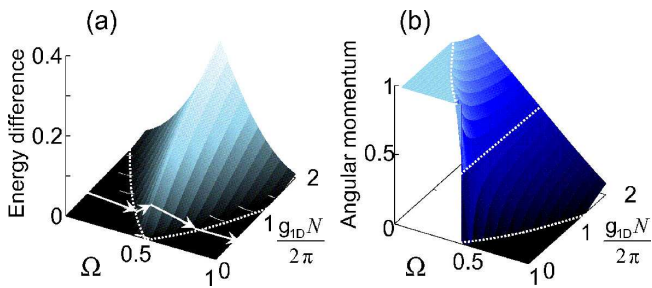}
\caption{
(a) Energy difference between the metastable plane-wave and soliton states.
The soliton solutions exist in the area sorrounded by two phase boundaries (white dotted curves).
(b) Corresponding angular momentum.
The soliton solutions make it possible
to smoothly connect quantized integer values of average angular momentum.
}
\label{fig2}
\end{figure}

Bifurcation of the soliton train from the plane wave constitutes
a second-order quantum phase transition with respect to $g_{\rm 1D}$ and/or $\Omega$.
Figure~\ref{fig2}(a) shows the energy difference $E_J^{\rm (sol)}-E_J^{\rm (pw)}$ 
between the two solutions
\begin{eqnarray}
E_J^{\rm (pw)}&=&(\Omega-J)^2+g_{\rm 1D}N/(4\pi), \nonumber\\
E_J^{\rm (sol)}&=&g_{\rm 1D}N/(2\pi)+\left[3KE-K^2(2-k^2)\right]/\pi^2\nonumber\\
\!+4K^2&&\!\!\!\!\!\!\!\!\!\!\!
\left[3E^2\!-\!2(2-k^2)KE\!+\!K^2(1\!-\!k^2)\right]\!/\!(3\pi^3g_{\rm 1D}N).\nonumber
\end{eqnarray}
This kind of bifurcation does not occur from the ground-state energy.
However, for metastable states a bifurcation can occur between the plane-wave
state and the soliton state with the same winding number $J$.
After bifurcation, the soliton energy $E_J^{\rm (sol)}$ becomes
larger than $E_J^{\rm (pw)}$.
Furthermore, at $\Omega=0.5$, $E_0^{\rm (sol)}$ and $E_1^{\rm (sol)}$
are degenerate with a $\pm\pi$-phase jump in the condensate wave function.
The derivatives of the energies $\partial E^{\rm (sol)}_{J}/\partial \Omega$ 
and $\partial E^{\rm (pw)}_{J}/\partial \Omega$ have a kink at
the boundary as can be verified analytically. This identifies
the second-order quantum phase transition~\cite{QPT}, which
occurs along a curve in the $\Omega$-$g_{\rm 1D}$ plane.

Figures~\ref{fig1} (i)-(vi) illustrate a continuous change in 
the topology along a higher-energy, soliton path shown in Fig.~\ref{fig2} (a) 
with white arrows.
Following this path in Fig.~\ref{fig1}, as $\Omega$ increases starting
from (i) the plane wave with $J=1$,
(ii) solitons start to form past a critical point $\Omega_{\rm cr}$.
(iii) The density notch deepens for $\Omega_{\rm cr} \le \Omega \le 0.5$.
At $\Omega=0.5$ it forms a node, the phase of the soliton jumps by $\pi$,
and the energies of the solitons with phase winding number 1 and 0
are degenerate. (iv), (v) The soliton with phase winding $J=0$ deforms continuously
as $\Omega$ increases. (vi) Finally, the state goes back to
the plane-wave state with phase winding $J=0$~\cite{lower}.

The angular momentum $L/N=\int d\theta \psi^*(-i\partial_{\theta})\psi$
of the metastable states changes continuously
along the soliton path. For the plane wave state, $L^{\rm (pw)}_{J}/N=J$ is quantized;
in contrast, for the soliton
$L^{\rm (sol)}_{J}/N=\Omega +{\cal S} \sqrt{2 f_{\rm sn}g_{\rm sn}h_{\rm sn}}/(g_{\rm 1D}N\pi^2)$
is non-integer, as shown in Fig.~\ref{fig2}(b).  Thus a continuous
change of angular momentum is possible for 1D Bose systems
by taking the metastable states with energy slightly higher than
that of the ground state. 

We next investigate the stability of the metastable states using Boguliubov
theory~\cite{01:FS,00:GACZ}, and identify the curve in the $\Omega$-$g_{\rm 1D}$ plane where
the soliton solutions bifurcate from the plane-wave solutions.
A stationary solution $\psi$ of the GPE subject to a small perturbation
$\delta$ evolves in time as
$\tilde{\psi}(t)=e^{-i\mu t}[\psi+\sum_n (\delta u_n e^{-i\lambda_nt}+\delta v_n^* e^{i\lambda_n^* t})]$,
where $(u_n, v_n)$ and $\lambda_n$ are eigenstates and eigenvalues of
the Boguliubov-de Gennes equations (BdGE), and $n$ denotes the index of the eigenvalues.

For the plane-wave state with phase winding $J$,
the eigenvalues of the BdGE are obtained as
$\lambda_n^{(J,{\rm pw})}=[n^2( n^2+g_{\rm 1D}N/\pi )]^{1/2}-2n(\Omega-J)$.
Then $\lambda_{-1}^{(J=1, {\rm pw})}$ is negative, 
monotonically increases for $\Omega \in [0,\Omega_{\rm cr})$, 
and crosses zero at 
$(g_{\rm 1D}N)_{\rm cr}/(2\pi) - 2(\Omega_{\rm cr}-J)^2+1/2=0$. 
Thus the 
metastable state $\psi=e^{i\theta}/\sqrt{2\pi}$ is
{\it thermodynamically} unstable.
The plane-wave limit of the soliton solutions $\eta \to 0$ occurs when
$(g_{\rm 1D}N,\Omega)$ approach their critical values from above.
We also obtained the eigenvalues of the BdGE when $\psi$ is taken 
as a soliton. 
The eigenstates obtained from a soliton involve the Nambu-Goldstone mode
$\lambda_{\rm NG}^{(J=1, {\rm sol})}$, i.e., the zero-energy rotation mode, 
associated with the rotational symmetry breaking of the soliton solution.
At the critical values of $(g_{\rm 1D}N,\Omega)$,
$\lambda_{-1}^{(J=1, {\rm pw})}=\lambda_{\rm NG}^{(J=1, {\rm sol})}$
and other eigenvalues $\lambda^{(J, {\rm sol})}$ are nearly degenerate with 
$\lambda^{(J, {\rm pw})}$.
There is no negative Boguliubov mode in the soliton regime.  Thus
the soliton state is linearly stable.


Finally, we investigate how continuous change in the
angular momentum per particle observed in the mean field theory is
described in terms of quantum many-body theory. Lieb derived
two kinds of excitation branches in the thermodynamic limit,
``Type I'' and ``Type II'' excitation branches of Eq.~(\ref{Hamiltonian}) 
with $\Omega=0$~\cite{63:LL}.  The Type II excitation was shown to 
be a dark soliton branch~\cite{80:IT}. 
We also determined low-lying excitation energies in a finite-size 
waveguide in a wide range of $g_{1\mathrm{D}}N$ via the Bethe
ansatz~\cite{04:JB,07:SDD,footnote3}.

\begin{figure}[t]
\includegraphics[scale=0.48]{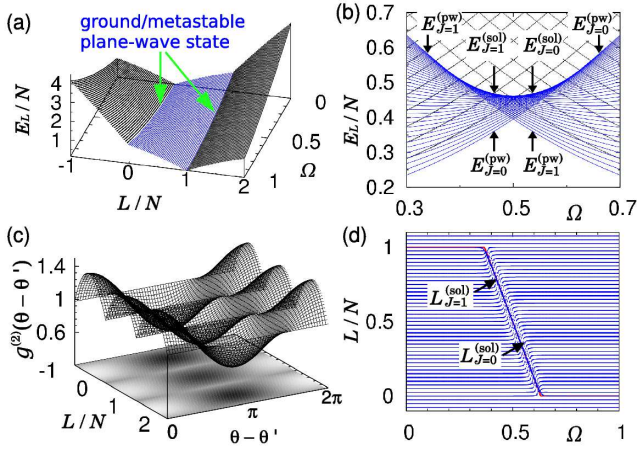}
\caption{
(a) The lowest energy of the Hamiltonian for each angular momentum subspace
for $N=40$, $g_{\rm 1D}=2\pi\times 7.5 \times 10^{-3}$.
(b) Enlargement of (a) near the critical point. Densely packed blue curves 
are spectra within $L \in \{0,N\}$. Otherwise the spectra are sparse.
(c) Two-body correlation function of eigenstates for each angular momentum subspace.
(d) Expectation value of the angular momentum of each eigenstate, in the presence of 
symmetry breaking potential $\hat{V}$ as a function of $\Omega$. 
}
\label{fig3}
\end{figure}

In order to study the metastable states
we need to investigate the $N^{\mathrm{th}}$ order
highly excited states. 
We diagonalize Eq.~(\ref{Hamiltonian}) in a basis which
provides equivalent results to those obtained with the Bethe
ansatz~\cite{footnote4} for the purposes of physical insight. 
The basis is taken as $|n_{-1},n_0,n_1,n_2\rangle$, subject to
conditions $\sum_l n_l = N$ and $\sum_l ln_l = L$, where
$n_{l}$ is the number of atoms with single-particle angular
momentum $l$ and $L\in \{-N,\dots,2N\}$ is the total angular
momentum. 
Figure~\ref{fig3}(a) shows the energies $E_L \equiv \langle
L,N|\hat{H}|L,N\rangle$ of yrast states $|L,N \rangle$,
i.e., the lowest-energy state for a fixed value of $L$ and $N$.
The index $N$ will be dropped, as it is fixed. The
curvature of the surface is independent of $N$ but the density
of states with respect to $L/N$ increases as $N$ becomes
larger. Three kinds of states appear in this energy landscape:

(i) Ground state: The ground state of the Hamiltonian is
$|L=0\rangle$ with energy 
$E_{L=0}/N \simeq \Omega^2+g_{\rm 1D}N/(4\pi)$ for $\Omega \in [0,0.5)$, 
and $|L=N\rangle$ with energy $E_{L=N}/N \simeq (\Omega-1)^2+g_{\rm 1D}N/(4\pi)$ for
$\Omega \in [0.5,1)$, respectively.
These are in agreement with the mean-field ground states 
$\psi=1/\sqrt{2\pi}$ and $\psi = e^{i\theta}/\sqrt{2\pi}$, respectively.

(ii) Metastable plane-wave states:
Similarly, the many-body counterparts of the metastable plane-wave solutions
of the GP equation are $|L=N\rangle$ for $0\leq \Omega < 0.5$ and
$|L=0\rangle$ for $0.5 \leq \Omega < 1$, respectively.

(iii) Seeds of broken-symmetry states: All many-body
eigenstates are rotationally invariant because they respect the
symmetry of the Hamiltonian, i.e., there are {\it no}
broken-symmetry eigenstates. However, as shown in
Fig.~\ref{fig3} (b), the eigenvalues $\{E_{L=0}, \ldots,
E_{N}\}$ (densely packed blue curves) cross each other in between 
the ground and metastable plane-wave states, and $\{E_{L=1}, \ldots,
E_{N-1}\}$ become larger than the metastable plane-wave state 
energies within a certain range of $\Omega$. The regime of this
quasi-degenerate level crossing can be identified with the
soliton regime given by the mean-field theory, as we will show.
Furthermore, the envelope of the highest eigenvalues in the
quasidegenerate regime coincides with $E^{\rm (sol)}_{J=0, 1}$
in the limit $N\to \infty$. In the absence of interaction
($g_{\rm 1D}=0$), all the levels with $L \in \{0,N\}$ are
degenerate for $\Omega=0.5$ only. The degeneracy inherent
in a level crossing permits the existence of a broken symmetry
state; in our case, this is the soliton solution corresponding
to the mean-field theory prediction.

Figure~\ref{fig3} (c) shows the two-body correlation function
$g^{(2)}(\theta-\theta') \equiv
\langle\hat{\psi}^{\dagger}(\theta)\hat{\psi}^{\dagger}(\theta')
\hat{\psi}(\theta')\hat{\psi}(\theta)\rangle / [\langle
\hat{\psi}^{\dagger}(\theta)\hat{\psi}(\theta)\rangle$ $
\langle
\hat{\psi}^{\dagger}(\theta')\hat{\psi}(\theta')\rangle]$,
where the expectation value is taken with respect to 
each yrast state $|L\rangle$. The function $g^{(2)}$ is
independent of $\Omega$ and has a single peak. When $L/N$ is an
integer, $g^{(2)}=1$; when $L/N$ is half- integer,
$g^{(2)}$ strongly deviates from 1. The one-body
density $g^{(1)} \equiv \langle
\hat{\psi}^{\dagger}(\theta)\hat{\psi}(\theta)\rangle=2\pi/N$
is a constant for any eigenstate, since the system is
rotationally invariant.

In order to force {\it quantum solitons} 
to appear in $g^{(1)}$, we add a symmetry-breaking perturbation of the form
$\hat{V}=\varepsilon\sum_{l\in\{-1,0,1,2\}}
(\hat{c}_{l+1}^{\dagger}\hat{c}_l+{\rm h.c.})$, $\varepsilon
\ll g_{\rm 1D}$. The total angular momentum $L$ is no longer a
good quantum number, and the eigenvalue problem is thus given
in a general form,
$(\hat{H}+\hat{V})|\Psi_n\rangle=E_n|\Psi_n\rangle$ where $n
\in \{0,1,\dots\}$ is the energy eigenvalue index. 
Employing the basis $|L\rangle$, where $L \in \{-N,2N\}$, 
the average angular momentum per particle is shown
in Fig.~\ref{fig3}(d) for each eigenstate 
for $\varepsilon = 5\times 10^{-3}$.
Outside the soliton regime, angular momenta remain integer values unaffected
by the small perturbation. In the soliton regime, on the other hand, 
by a superposition of eigensates $|L\rangle$ in the absence of the perturbation 
some of the angular momenta converge to a noninteger value,
which indeed agrees with $L^{\rm (sol)}_{J=0,1}$. We have also calculated
the one-body correlation function $g^{(1)}$
for all eigenstates in the presence of the symmetry breaking perturbation, 
and confirm that $g^{(1)}$ has a single density notch in the soliton regime
with the depth close to the mean-field grey soliton.


In conclusion, we found a denumerably infinite set of paths to connect plane-wave states
via soliton trains in a metastable system of scalar bosons on a ring.  Associated with
this transition, the energy of the solitons bifurcates, and a continuous change
in the angular momentum becomes possible in the mean-field theory. We made
a link between these mean-field results and the full quantum theory by showing that
quasidegenerate energy levels are related to the formation of quantum solitons.


We thank Joachim Brand and Yvan Castin for useful discussions.
This material is based upon work supported by the National
Science Foundation under Grant No. PHY-0547845 as part of 
the NSF-CAREER program, and a Grant-in-Aid for Scientific Research 
(Grant NO. 17071005).



\begin{thebibliography}{50}
\bibitem{footnote1}
The Hamiltonian includes a constant term proportional to $\Omega^2$ which is
associated with rigid-body rotation and only shifts the origin of the total
energy. 
The many-body wave functions
$\Psi_{\rm rot}(\{\theta\})$ of
this Hamiltonian are obtained by the transformation of
the solution $\Psi_{\rm lab}(\{\theta\})$
in the rest frame (original Lieb-Liniger Hamiltonian),
via $\Psi_{\rm rot}(\{\theta\})
=\prod_j \exp(2i\Omega\theta_j)\Psi_{\rm lab}(\{\theta\})$.
\bibitem{k}
$2\pi \Omega {\cal S}=
2(1-k^2) (jK)^2 2 f_{\rm sn}/(g_{\rm sn}h_{\rm sn})
+\sqrt{2f_{\rm sn}h_{\rm sn}/g_{\rm sn}}
+j\pi\left[1-\Lambda_0(\epsilon \backslash \alpha)\right]$,
where $\Lambda_0$ is Heuman's lambda function, $\alpha=\arcsin k$,
and $\epsilon \equiv \arcsin \left(f_{\rm sn}/h_{\rm sn}\right)$.
\bibitem{lower}
We note that the amplitude remains uniform, and the phase looks like Fig.~\ref{fig1}(i)
for $\Omega\in[0,0.5)$, and Fig.~\ref{fig1}(vi) for $\Omega \in[0.5,1)$ with a
$\pi$-phase jump at $\Omega=0.5$, if one takes the lower energy path.
\bibitem{footnote4}
For our interaction regime the convergence of this truncation
is confirmed by the comparison with the finite-size Bethe-ansatz method.
\end{thebibliography}
\end{document}